IAC-10-C4.8.7

# PROGRESS IN REVOLUTIONARY PROPULSION PHYSICS

**M. G. Millis**
Tau Zero Foundation, Fairview Park, United States, marc@tauzero.aero

Prior to 1988, traversable wormholes were just science fiction. Prior to 1994, warp drives were just fiction. Since then, these notions matured into published scientific discourse, where key issues and unknowns continue to be raised and investigated. In 2009, the American Institute of Aeronautics and Astronautics published a peer-reviewed, expansive technical volume on these and other investigations toward breakthrough propulsion. This paper summarizes the key assertions from that 739-page volume, describing the collective state-of-the-art and candidate research steps that will lead to discovering if, or how, such breakthroughs might finally be achieved. Coverage includes: prerequisites for space drive physics, manipulating gravity or inertia for propulsion, lessons from superconductor experiments, null results with "lifters", implications of photon momentum in media, quantum vacuum physics, and the faster-than-light implications of general relativity and quantum non-locality.

## I. INTRODUCTION

The goal of timely interstellar flight – to reach other habitable worlds within a human lifespan – cannot be achieved with even the most refined technological applications of accrued physics. The exhaust velocity and propellant mass required when applying the rocket equation, or the power level required for photon momentum transfer, are so high as to fall into the realm of the seemingly impossible [1: Ch 2].

To circumvent these limits, it is desired that new, advantageous, propulsion physics awaits discovery. For example, if it were possible to move a spacecraft using the interactions between the craft and its surrounding space without needing propellant (i.e. a "space drive") then the energy requirements would drop from exponential to squared functions of trip velocity [1:145]. If faster-than-light travel becomes possible, then the light-years spanning star systems become traversable within a human lifespan.

Objectively, such desired breakthroughs might turn out to be impossible, but progress is not made by conceding defeat. In 2009, the first scholarly book was published that examines the correlations between these desired breakthroughs and contemporary physics. This book, *Frontiers of Propulsion Science* (*FPS*), is the primary reference upon which this paper is based [1]. This paper presents a condensed summary of the approaches from that book and identifies next-step questions toward determining if, and how, such breakthroughs might eventually be achieved. Also, suggestions are offered for navigating amongst the uncertainties of such provocative, nascent research.

This research falls within the realm of *physics* rather than *technology*, with the distinction that physics is about uncovering the laws of nature while technology is about applying that science to build useful devices. Before the technology of such devices can be engineered, new physical principles must first be discovered, confirmed, and modelled.

## II. COLLECTIVE STATE-OF-THE-ART

Although *warp drives* and *wormholes* might sound like science fiction, investigations into them are appearing in increasing numbers in professional journals. Concepts of space drives and notions of manipulating gravitational or inertial forces are also entering professional discourse.

Although no breakthroughs appear imminent, the subject has matured to where the relevant questions have been broached and are beginning to be answered. While it is too soon to predict if, or when, the breakthroughs will be found, what is certain is these goals are finally approachable through rigorous research.

Dozens of concepts have been introduced that span stages 1 through 3 of the scientific method; that is, *defining the problem*, *collecting data*, and *articulating hypotheses*. Some have matured to stage 4, *testing hypotheses*, but most of the completed tests have revealed misinterpretations of previously known, unremarkable effects.

Recent advances in general physics, such as attempts to decipher *Dark Matter*, *Dark Energy*, *quantum vacuum energy*, and other phenomena, also fuel progress toward solving the key issues and unknowns. That progress, however, is often cast in the context of cosmological curiosities rather than the utilitarian motivations of spaceflight.

While general science continues to assess cosmological data regarding its implications for the birth and fate of the universe, a spaceflight focus will cast these observations in different contexts, offering insights that might otherwise be overlooked from the curiosity-driven inquiries alone. Homework problems to help teach general relativity now include warp drives and traversable wormholes [2: 489]. Even if there are no spaceflight breakthroughs to be found, adding the inquiry of spaceflight expands our ability to decipher the lingering mysteries of the universe.

### III. NOVICE ORIENTATION

It is understandably difficult for the non-expert to gauge the prospects of such revolutionary pursuits quickly. Revolutionary ideas, by their very nature, break from the familiar and can look initially as nonsensical as genuinely errant ideas. Distinguishing these is easy in retrospect. The errant ideas fade away, while the viable ideas survive, often with infamous dismissive quotes such as: "Space travel is utter bilge" (uttered by Dr. Richard van der Riet Wooley, one year before Sputnik, 1957).

Realizing this difficulty, *FPS* also examined lessons from prior revolutionary work to provide suggestions for how to navigate productively amongst this uncertainty [1: Ch 22]. To aid the reader, a condensed version of those suggestions follows.

#### Culling Progress

To avoid the extremes of reflexive dismissals and sensationalist hype common with revolutionary pursuits, it is recommended to focus on the *rigor* and *objectivity* of the concepts rather than trying to judge their *feasibility*. An impartial feasibility assessment on unfamiliar topics is as difficult as a research task unto itself. Instead, the level of rigor is easier to judge.

Classic symptoms of non-rigorous work are reflected in Langmuir "pathological science" [3], Sagan's "baloney detector" [4], Baez's "Crackpot Index" [5], and the lessons from the NASA Breakthrough Propulsion Physics Project [1: Ch 22]. Representative symptoms from those sources include:
- Selectively addressing supporting evidence while neglecting contrary evidence or the possibility of false-positives.
- The magnitude of effect remains close to the limit of detectability, along with claims of great accuracy.
- Drawing conclusions from inadequate sample sizes (Statistics of small numbers).
- Confusing *correlation* with *causation*.
- Lack of relevant reference citations.

In addition to the easy-to-spot symptoms of non-rigorous work, attributes that indicate rigorous work include:
- The submitter is aware of the focal make-break issues related to their approach.
- The submitter is cognizant of much of the reliable relevant literature.
- Any alternative or unconventional interpretations of known phenomenon are accompanied by correct reference citations of those phenomena.

An additional tactic to make progress is to define success in terms of gaining reliable knowledge rather than achieving a breakthrough. This shifts attention away from the temptation to oversell claims and instead focuses on the rigor and impartiality behind the assertions and findings. This also allows failed approaches to become valuable lessons to guide future decisions.

#### Identifying Critical Issues and Unknowns

To identify the focal research questions, the desired propulsion goals are contrasted to the accrued physics. Next-step research opportunities lay at the intersection between the issues evoked by the propulsion ideas and unresolved problems in contemporary physics.

A more systematic and detailed version of this process is John Anderson's *Horizon Mission Methodology* [6]. This process was applied by the NASA Breakthrough Propulsion Physics Project and the lessons learned are summarized in the last chapter of *FPS* [Ch 22].

Additionally, this process can be graphically plotted as a map, where the foundational physics borders one side and the propulsion goals the other. Branching out from each toward the middle are the more specific unresolved issues. Where the issues and unknowns intersect – those intersections define areas of needed research. A version of such a map, taken from page *xxiv* of *FPS*, is presented in Figure-1, but its reproduction here does not provide the resolution to distinguish its full details. It does, however reflect the principle behind such maps. Further details are offered in *FPS* [p.695-697].

Another succinct way to convey this process is to acknowledge the *objections* encountered when contemplating propulsion breakthroughs, and then realize that these *objections* actually suggest research *objectives*.

For example, the notion of thrusting without propellant evokes objections of violating conservation of momentum. This, in turn, suggests that space drive research *must* address conservation of momentum. From there it is found that many relevant unknowns still linger regarding the source of the inertial frames against which conservation is referenced. Therefore, research should revisit the unfinished physics of inertial frames, but in the context of propulsive interactions.

### IV. NOTIONAL PROPULSION APPROACHES

Over three-dozen concepts toward breakthrough prolusion and power are itemized in Table-1. None of these are at the stage of a confirmed working device, but all are useful in identifying relevant questions needing deeper study. Absent a taxonomy for this nascent topic, these have been categorized first by the grand challenge they address, and then by their method or physics discipline.

The *grand challenges* used to group these are based on the key barriers to practical interstellar flight and were the guiding goals of the NASA Breakthrough Propulsion Physics Project: 1) non-propellant propulsion, 2) faster-than-light travel, and 3) energy breakthroughs related to those two goals.

The table provides: a name for each approach that indicates its mechanism; an author name for beginning literature searches; the page from the *FPS* book where the topic discussion begins; a classification of the work as *notional*, *theoretical, experimental*, or an *engineering concept*; a classification of its progress via the scientific method; and an abbreviated assessment of key issues.

Although the details are too expansive to include here, the main recurring themes include: issues of space drive physics, approaches to manipulating gravity or inertia, lessons from superconductor experiments, null results with "lifters", implications of photon momentum in media, quantum vacuum physics, and the faster-than-light implications of general relativity and quantum non-locality.

Space Drive Physics

*Space drive* is a generic term to encompass the notions of using the interactions between the vehicle and its surrounding space to induce motion. The motivating goal is to eliminate the need for propellant. As a consequence, the energy requirements drop from being exponential functions of trip velocity to only squared functions. The primary issues, regardless of embodiment, are *conservation of momentum*, and *net external thrust*.

Conservation of momentum

Thrusting requires a reaction mass as inferred by Newton's 3$^{rd}$ law: "*for every action there is an equal and opposite reaction*." As the craft moves forward, the reaction mass must move rearward such that total momentum of the system is conserved. Absent *obvious* matter in space to use as a reaction mass, space drive concepts must delve into the unknowns about the sources of inertial frames, quantum vacuum energy, or the unresolved physics of photon momentum in media.

Despite their ubiquitous nature, inertial frames are still not fully understood. That they exist is certain, but what *causes* them to exist – and if these sources could constitute a reaction-media – are still unknown. Multiple versions of "*Mach's principle*" exist in the literature (speculates that inertial frames are due to surrounding matter), where some have been dismissed, but others remain unresolved curiosities [7].

Even estimating the mass density of spacetime is contentious. Depending on whether one uses the estimates of visible and dark matter, or inferences of the stress-energy of spacetime, resulting estimates can span from the trivial value of $10^{-26}$ kg/m$^3$ to the enormous value of $10^{25}$ kg/m$^3$ [1: 131].

Quantum vacuum energy is the next consideration for a propulsive reaction-media due in part to its estimated equivalent mass density and its incomplete understanding in physics. Even though the notion of zero-point energy dates back to 1911, and the first experimental evidence occurred in 1924 (Mulliken's boron monoxide spectroscopy) [1: 575], the physics of vacuum energy continues to be actively explored [8]. Depending on factors chosen for the calculation, the equivalent mass density of the quantum vacuum energy can span from the insignificant $10^{-26}$ kg/m$^3$ to the enormous value of $10^{98}$ kg/m$^3$ [1:131].

Net External Thrust

Another major issue for space drives is ensuring that the propulsive effect actually moves the vehicle relative to the surrounding space rather than just inducing forces *internal* to the vehicle. A common mistake, especially with electromagnetic schemes, is to have thrusting methods where the reaction mass is part of the vehicle itself. This is analogous to pushing on a dashboard from inside a car in an attempt to move the car.

Space Drives Compared to Warp Drives

The category of *space drives* is distinct from the notion of *warp drives* and *wormholes*, in that space drives interact *with* spacetime instead of *warping* spacetime. Warp drives and wormholes are rooted in the Riemannian geometry of general relativity, where sufficient energy densities can warp spacetime analogously to how a huge gravitational mass bends spacetime. In contrast, most space drive concepts begin with Newtonian representations where the operative goal is to interact with reaction mass imbedded in the properties of spacetime. This also implies, therefore, that space drive concepts will be light-speed-limited since they operate *within* spacetime.

To convey this in terms of an analogy, consider moving an automobile across a landscape. General relativity (space warping) allows us to consider how to reshape or move sections of the landscape so that the automobile (and everything in the vicinity) will roll passively downhill toward the desired destination. This requires substantial energy expenditures, but also opens the way for creating faster-than-light pathways. The Newtonian perspectives, on the other hand (space drives), consider how to move the automobile under its own power relative to that landscape, analogously to tires pushing against the ground.

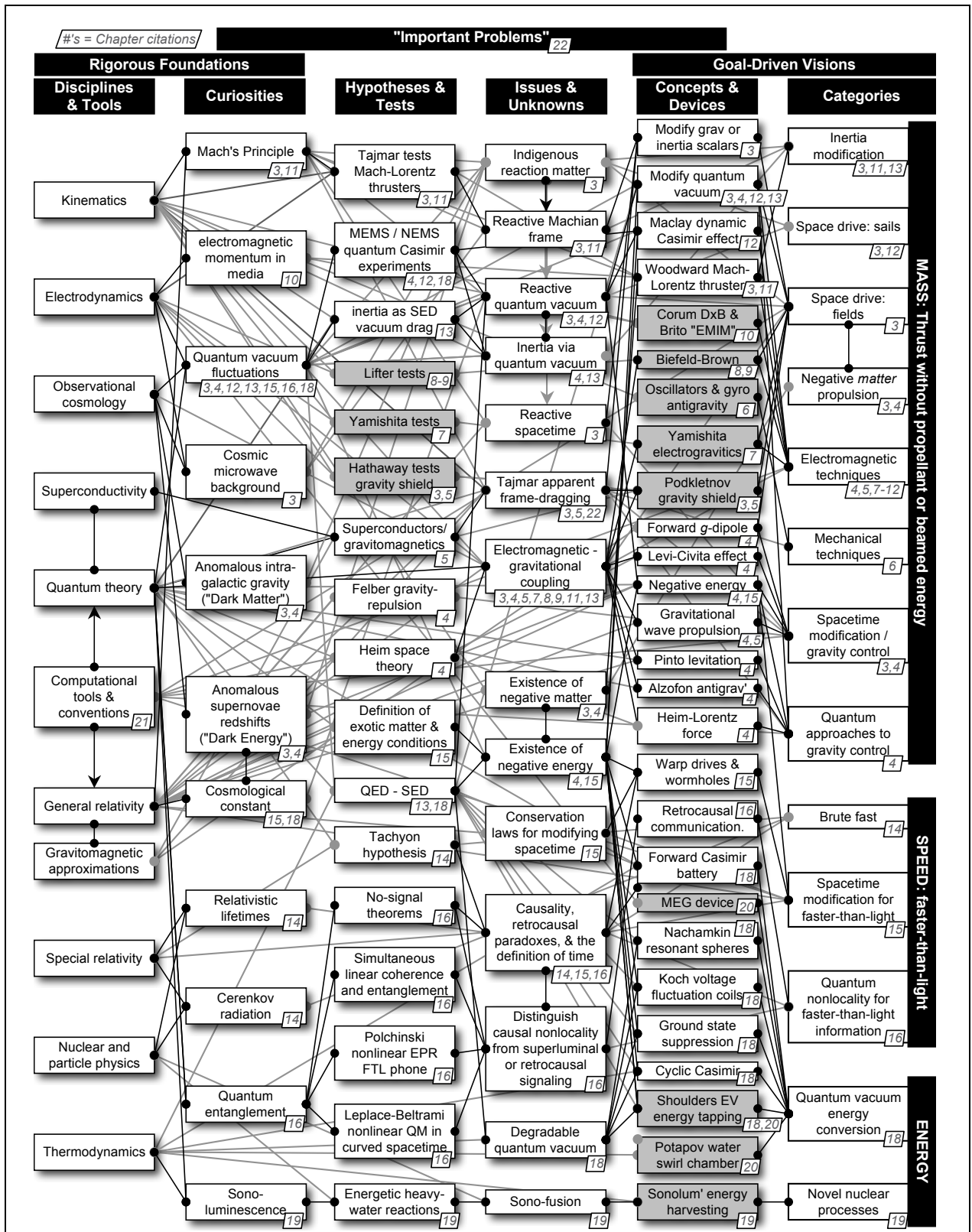

**Fig. 1: Relational Depiction of the Contents of *Frontiers of Propulsion Science*.** The far left column lists relevant disciplines. The far right column shows the goals broken down into categories of different approaches. Columns in between are where the various unknowns, critical issues, and curious effects await resolution. Shaded blocks indicate nonviable approaches.

| Investigation | Author Search Suggestions | *FPS* page | Type* & Sci** | | Key Issue or Next Research Step |
|---|---|---|---|---|---|
| **PROPELLANTLESS PROPULSION APPROACHES** | | | | | |
| *General Approaches* | | | | | |
| Mechanical oscillation thrusters | – | 249 | C | 4 | **Error**: Misinterpretation of observations |
| Mechanical gyroscopic thrusters | Laithwaite, E. | 254 | C | 4 | **Error**: Misinterpretation of observations |
| Asymmetrical capacitors ("Lifters") | Brown, T.T. | Ch 8-9 | C | 4 | **Error**: Misinterpretation of observations |
| Sails: *Differential, Induction, & Diode* | Millis, M. G. | 152 | N | 1 | Passive sails violate conservation laws, active sail (*induction*) only notional idea |
| Field drive, Diametric (negative mass) | Bondi, Forward | 160, 180 | T | 3 | Requires negative inertia |
| Field drive, Disjunction | Millis, M. G. | 162 | N | 1 | Requires separable properties of mass |
| Field drive, Induced gradient | Millis, M. G. | 164 | N | 1 | Requires connection to reaction media |
| Field drive, Bias drive (inertial frame) | Millis, M. G. | 165 | N | 1 | Inertial frame physics incomplete |
| Anomalous frame dragging observed | Tajmar, M. | 243 | E | 2 | Inexplicable data, not confirmed |
| Enhanced photon momentum (media) | Corum, Brito | Ch 10 | T, E | 4 | **Error**: Misleading mix of conventions |
| Vibrating mirror Casimir drive | Maclay/Forward | Ch 12 | T | 3 | Requires improved conversion efficiency |
| *Contemplating Inertial Manipulations* | | | | | |
| Inertia modified rocket or launch pad | – | 138 | N | 1 | Notional foundations (some misleading) |
| Inertial oscillation thruster | Woodward | 156 | T | 3 | Momentum transferred to inertial frame |
| | | Ch 11 | E | 4 | Experimental data still inconclusive |
| Inertia as quantum vacuum drag force | Haisch, et al | Ch 13 | T | 3 | Speculation open, but some in error |
| *Contemplating Gravitational Manipulations* | | | | | |
| Dipole torus field generator | Forward, R.L. | 185 | T | 3 | Enormous engineering required |
| Gravity-like solenoid stress energy | Levi-Civita | 198 | T | 3 | Enormous engineering required |
| Gravitational wave rocket | – | 201 | T | 3 | Non-viable thrusting conversion |
| Gravitational wave transducer | Chiao, R. | 242 | T | 3 | Conversion efficiency limits |
| Quantum corrections | Forward, et al | 211 | T | 3 | Physics still developing |
| Quantum fluctuations and gravity | Calloni, et al | 213 | T | 1 | Speculations open, but some in error |
| Nonretarded interatomic dispersion | Pinto, F. | 215 | T | 3 | Physics still developing |
| Gravitophoton perspective | Dröscher, Heim | 218 | T | 1 | Not yet rigorously articulated |
| Modify mass via nuclear entropy | Alzofon | 221 | T | 1 | Speculations require more definition |
| Superconductor gravitational shield | Podkletnov | 140, 239 | E | 4 | **Error**: Not rigorous, not repeatable |
| Superconductor discharge force beam | Podkletnov | 242 | E | 4 | **Error**: Not rigorous, not repeatable |
| Electro-gravitational patent | Yamishita | Ch 7 | E | 4 | Data not above statistical significance |
| **FASTER-THAN-LIGHT (FTL) APPROACHES** | | | | | |
| Traversable wormholes | Visser, M. | 484 | T | 3 | Requires enormous negative energy |
| Warp Drives | Alcubierre, M. | 487 | T | 3 | Far less energy efficient than wormholes |
| Quantum non-locality | Cramer J., et al | Ch 16 | E, T | 3 | Physics still developing |
| Hyperspace | Schutz, B. | 485 | T | 3 | Not the FTL science fiction hyperspace |
| **ENERGY CONVERSION APPROACHES** | | | | | |
| *Quantum Vacuum Energy Approaches* | | | | | |
| Casimir coil battery | Forward, R. L. | 571 | C | 3 | Feasible as one-shot, efficiency limits |
| Resonant dielectric spheres | Mead/Nachamkin | 572 | T | 3 | Extraction/conversion scheme pending |
| Voltage fluctuations in coils | Koch, Blanco | 577 | E, T | 3 | Extraction/conversion uncertain |
| Ground state reduction via Casimir | Puthoff/Haisch | 579 | E | 3 | Evokes unresolved plenum issues |
| Tunable Casimir cavities | Puthoff/Iannuzzi | 583 | T | 3 | Net energy from cycle & plenum issues |
| *Misc Approaches* | | | | | |
| Electromagnetic Vortex evidence | Shoulders, K. | 585 | E | 2 | Energy conversions undeveloped |
| Motionless Electromagnetic Generator | Bearden, T. | 644 | E | 4 | **Error**: Misinterpretation of observations |
| Sonoluminescence anomalies in $^2H_2O$ | Wrbanek, et al | Ch 19 | E | 2 | Observed anomaly, not yet understood |
| Low Energy Nuclear Reactions | Storms, E | 645 | E | 2 | Anomalies not yet consistently observed |

**Table 1: Examples of Concepts Toward Breakthrough Propulsion and Power**

\* Type abbreviation key:
 **C** = Engineering concept, **E** = Experiments, **N** = Notional material for thought experiments, **T** = Theoretical.

\*\* Scientific method level abbreviation key:
 **0** = Pre-science, **1** = Problem Defined, **2** = Collecting Data, **3** = Hypothesis Articulated, **4** = Hypothesis Tested.

Controlling Gravity or Inertia For Propulsion

A recurring theme to breakthrough propulsion is to consider that gravitational and inertial forces can be modified. In addition to the space drive issues just discussed, these approaches also face the challenge of exploiting as-yet-unresolved correlations between the fundamental forces.

Contemporary physics has not yet fully resolved the connections between gravitation, electromagnetism, and the other fundamental forces. Although general relativity provides models that are accurate on the scale of solar systems, the anomalous observations leading to the hypotheses for Dark Matter and Dark Energy, plus the anomalous trajectories of deep space probes [9, 10], indicates that there is more physics yet to be discovered. Additionally, quantum physics – whose models of electromagnetic behaviour are profoundly useful on extremely small scales – has not been successfully extended to cosmological scales.

Approaches that rigorously explore these unfinished areas, such as the discipline of *quantum gravity*, are numerous.

Woodward's Inertial Oscillation Thrusters

One approach dating back to 1990 is based on the perspective that inertia is related to energy density and interactions with inertial frames [11]. Theoretically Woodward's thrusting concept satisfies conservation of momentum by evoking a literal interpretation of *Mach's principle*, where the surrounding mass of the universe is the reaction mass. While experiments are underway, the results are still inconclusive. One series of tests are documented in *FPS* [Ch 11].

Despite this uncertainty, the issues regarding the role played by an inertial frame and internal energy lead to relevant thought experiments. One example is the displacement of a system's center of mass, graphically illustrated on page 158 of *FPS*. Such notions beg deeper inquiries about the reference frames against which inertial reference frames are measured.

Lessons From Superconductor Experiments.

In 1992, controversial claims of a "gravity shielding" effect using spinning superconductors were published in *Physica C* [12]. Subsequent independent testing was not able to replicate the claims, even with the participation of the lead researcher behind the original claims and having significantly more sensitive instrumentation (factor of 50 improvement) [13]. Because of the numerous investigations that this claim spawned, a chapter in *FPS* is dedicated to describing the experimental pitfalls to avoid when conducting such experiments [Ch 5].

More rigorous experiments by Martin Tajmar involving rotating cryogenics have reported anomalous inertial frame dragging [14] that has yet to be independently confirmed or reliably attributed to experimental misinterpretations. Given the propulsion relevance of inertial frames, this anomaly remains an unresolved and relevant curiosity.

At the same time that the anomalous laboratory frame-dragging was reported by Tajmar, the preliminary results from the *Gravity Probe B* mission showed similar anomalous torques on their gyroscopes. In the final report, however, these anomalies were attributed to "electrostatic interaction between the *patch effect* fields on the surfaces of the rotor and the housing" [15]. These torques, once modelled, were removed from the data on relativistic drift rate.

Null Results With "Lifters",

Two chapters in the *FPS* book are dedicated to assessing the sensationalistic claims behind "lifters," which are also referred to as "asymmetrical capacitor thrusters," and the "Biefeld-Brown effect." Both chapters, each following different investigative procedures, came to the same conclusion that the breakthrough claims are misinterpretation of ion wind.

Implications of Photon Momentum In Media

While photon momentum in vacuum is well understood, there are two different formalisms, the Minkowski and Abraham representations, to describe photon momentum in dielectric media. This has led to propulsion concepts where it appears that forces greater than photon momentum can be created by applying specific electromagnetic fields onto dielectric media.

Chapter 10 in *FPS* shows that the thrusting claims are misinterpretations from mixing these formalisms. When analyzing a system that involves photon momentum in media, such analyses should be strictly constrained to just using one of those formalisms. It is not clear which of these formalisms most closely matches nature, however, and this remains an unresolved issue in general physics.

Quantum Vacuum

The most succinct explanation for quantum vacuum energy links it to the *uncertainty principle*, where any system of harmonic oscillators can never be at an absolute zero energy state. More than just a theoretical notion, physical evidence exists to support the perspective that energy remains in a space even in the absence of all other energy sources.

Numerous energy-conversion concepts and a few propulsion concepts speculate on how to tap into this energy, with some more rigorous than others. It has been theoretically shown that a net thrust can indeed be induced by a "dynamical Casimir effect," but this initial embodiment is more feeble than a photon rocket [1: Ch 12].

Energy conversion is also possible in principle without violating thermodynamics, but these only function as one-shot devices that would require a subsequent energy to recharge them (i.e. Forward's Casimir Battery) [1: 571].

More generally, the notion of being able to continuously tap this energy is contentious, and centers on issues of whether the quantum vacuum can be viewed as a *plenum*, perhaps tied to other phenomena. Theoretical issues linger on which formalisms of quantum mechanics apply, the more refined Quantum Electrodynamics (QED), or Stochastical Quantum Electrodynamics (SED). Although multiple experimental approaches are proposed to investigate these, at the time of this writing none has been satisfactorily completed [1: 577-587]

Shortly after the completion of *FPS*, a scholarly book about the Casimir effect was published [8]. This provides an additional founding reference from which to explore this developing topic in physics.

Faster-Than-Light Implications of General Relativity

Based on the Riemannian geometry used in Einstein's general relativity, it has been theoretically shown that the light speed limit can be *circumvented* by manipulating spacetime itself. Rather than moving *through* spacetime beyond light speed, these approaches manipulate spacetime itself to either create shortcuts (wormholes) or to move 'bubbles' of spacetime (warp drives) faster than light can move *through* spacetime. The rate at which spacetime can move is inferred from the faster-than-light expansion of spacetime assumed to occur during the Big Bang.

The contentious issues are the implications of causal paradoxes (time travel), the magnitude of energy required, and the contention that the energy must be "negative." Although negative energy states are observed in nature, there are unresolved issues regarding the amount of negative energy allowed and for how long that energy can be used [1: 489-496].

Another conclusion reached is that warp drives are far less energy efficient than wormholes. Since a wormhole is a spacetime short cut, it already provides a faster-than-light passage once created. This holds for the smallest of wormholes that require the least energy. In contrast, the lowest energy warp drives are not faster-than-light. Warp drives become faster with more energy, but the energy demands are substantial. For example, around $10^{46}$ Joules of negative energy is required to create a faster-than-light wormhole of roughly 100m diameter, but if that same amount of energy is applied to an equal diameter warp bubble, the resulting speed is only 1% light speed [1: 491].

Another faster-than-light notion that merits mention is the science fiction idea of "hyperspace." This term means different things in fiction than in real general relativity. In the fictional context, hyperspace is as alternate realm without a light-speed limit. By travelling through *hyperspace* instead of *real* space, the fictional vehicle circumvents the light speed limit. In general relativity, however, a *hyperspace* or a *hypersurface* is a projection of a higher dimensional space onto a lower dimensional space. An analogy is the 2D surface (hypersuface) of a 3D globe. Reference [16] is a suitable textbook that describes the real science of hyperspace.

While theoretical foundations are substantial for faster-than-light spacetimes, laboratory exploration of these theories has not yet begun. Ultrahigh-intensity tabletop lasers might now be able to momentarily produce extreme electric and magnetic fields suitable for testing space-warping theory [1: Ch 15].

Faster-Than-Light in Quantum Non-Locality

In quantum physics there is the topic of *non-locality*, which includes the phenomenon of *entanglement*. More than one embodiment of entanglement exists, but for introductory purposes, consider two photons created at the same event. The paradoxical experimental finding is that these photons appear to be connected even when detected separately downstream of their origin. It appears somehow that this connection exceeds light speed. While the experiments are reliable, the unresolved issue is in the interpretation of the findings. This is an ongoing area of physics research.

To force the issue on the communication speed of this entanglement and the resulting causality-violating premise of faster-than-light communication, an experiment has been proposed to test the *'retro-causal'* signalling implications of such faster-than light assertions [1: Ch 16]. At the time of this writing, the experiment is being constructed by John Cramer, at the University of Washington, Seattle [17].

## V. UNFINISHED FOUNDATIONS OF PHYSICS

In addition to approaching the breakthroughs from the perspective of desired devices, one can also approach the goals by extending unfinished physics.

Physics continually seeks new knowledge to augment and correct errors in accrued knowledge. Over the last decades, observations have accumulated that are unexpected in terms of accrued models. A short list of these – which have relevance to propulsion and power – include:

- Cosmic Microwave Background Radiation provides an effective absolute reference frame for motion relative to the mean rest frame of the universe (measured by fore/aft Doppler shifts) [1: 132].
- Cosmic microwaves are anomalously homogeneous (once Doppler shifts are subtracted from the data).
- Origins for the inertial frame properties of space have not been determined.

- Anomalous space probe decelerations [9] and fly-by trajectories [10] are not explainable by known mechanisms.
- Anomalous intra-galactic binding led to the Dark Matter hypothesis and to theories for Modification of Newtonian Dynamics (MOND) [1:132].
- Anomalous galactic gravitational lensing suggests corroborating evidence toward the Dark Matter hypothesis.
- Anomalous supernovae red-shifts led to the Dark Energy hypothesis [1: 131, 192, 195]
- General relativity and quantum physics are still not compatible.
- The speed of gravitation, assumed to be identical to light speed, has not yet been measured.
- Gravitational forces have still not been explicitly and quantitatively linked to the other fundamental forces.
- The implications of the electroweak Higgs mechanism to the masses of subatomic particles remains uncertain, along with subsequent implications for macroscopic effects.
- Implications of quantum vacuum effects on cosmological scales remain unresolved.
- Normalization techniques to accommodate quantum vacuum energy divergences are still ad hoc and the physical implications of this energy remain unresolved.
- Interpretations of quantum mechanics, including entanglement, tunnelling, and non-locality remain unresolved.
- Effects on vacuum forces from cavity geometry, massive fields, dense matter, or motion have not been satisfactorily completed. Predicting many of these effects theoretically is beyond the calculation ability of QED.
- Experiments on "Low Energy Nuclear Reactions" have been sufficiently documented to indicate that genuine anomalies are present [18], but still lack rigorous uniformity of procedures and findings.

While physics attempts to resolve these unknowns in the most generally applicable manner, adding the implications of the desired propulsion and power goals provides an additional perspective from which to decipher these mysteries.

This was the approach taken in the 2002 study by the European Space Agency (ESA) when investigating breakthrough propulsion physics [19], which focused on these questions:
- Is there any evidence for deviations from the equivalence principle in orbit or during deep-spaceflight?
- Can the anomalous trajectories of deep space probes be better understood by launching a dedicated deep space probe?
- Can the gravitomagnetic effects being explored in ground based laboratories lead to new physical observations?

Regarding that last question, discretionary research by Martin Tajmar eventually led to observations of anomalous frame dragging near rotating cryogenics – work which still has yet to be independently confirmed or reliably attributed to other effects [14].

## VI. NEXT STEP RESEARCH

By contrasting the issues evoked from the approaches listed in Table 1 with the unfinished physics listed in the prior section, the following suggestions for next-step research are deduced. This is not a comprehensive list, but rather representative samples of the kind of research that could contribute to learning if, and how, spaceflight breakthroughs might be achieved.

Also, consider these suggestions to be general themes rather than specific tasks. For each of the bullets listed below, more than one research task could be proposed.
- Publish the results of null tests of claimed propulsion or power devices so the greater community can avoid re-investing in dead-end tests (e.g. Shawyer EM drive?).
- Modify existing symbolic mathematical computational tools to adopt preferred notational conventions for research toward propulsion methods. These conventions are detailed in Chapter 21 of *FPS*.
- Determine if measurable effects are possible using ultrahigh-intensity tabletop lasers to test the space-warping assertions from general relativity.
- Independently repeat or devise new experiments to explore Tajmar's unconfirmed observations of inertial frame dragging related to rotations of ultra-cold matter.
- Independently repeat or devise new experiments to test Woodward's inertia experiments – perhaps by determining if there are secondary implications that can be more definitively tested (data from the net-thrust tests are still inconclusive).
- Using emerging technology for micro- and nano-structure engineering, tangibly explore the physics of the quantum vacuum, exploring multiple geometries and materials, and examining whether external phenomena affect measurements. This includes influences of nearby dense matter, the orientation in a gravitational field, conjectured influences of diurnal or annual variations from the Earth's motion through the universe, etc.
- Explore vacuum energy experiments using negative index of refraction materials, ultra high electrical carrier density materials, and superconductors.

- Continue theoretical work on quantum forces in curved spacetime (i.e., gravitational field) with a focus on deriving effects that can be amplified to macroscopic levels.
- Complete Cramer's "retro-causality" entanglement experiments.
- Revisit prior theoretical attempts to mature *Mach's principle* into testable theories, but now including the following natural observations that were not known at the time of those earlier attempts: the absolute reference frame from the Cosmic Microwave Background Radiation, anomalous trajectories of deep space probes, and the anomalous observations that lead to the Dark Matter and Dark Energy hypotheses.
- Continue experiments that attempt to extract quantum vacuum energy, but with an emphasis on acquiring fundamental data rather than demonstrating new energy solutions (improves impartiality of findings).
- Search for violations of the equivalence principle in orbit or during deep-space coasting where any minuscule effects can integrate over time to become observable.
- Launch a deep space probe dedicated to, and simplified for, comparing its trajectory to the anomalous trajectories of other probes.
- Continue with theoretical explorations of wormholes and warp drives, focusing on minimal negative energy requirements and proceeding to consider the effects of finite rate-of-change for assembling the energy distributions required to activate wormholes or warp drives.
- Continue with theoretical explorations of wormholes and warp drives, focusing on momentum conservation on scales larger than the immediate scale of the wormhole or warp bubble (remote observers in asymptotically-flat spacetime).
- Proceed with experimental studies of anomalous heat phenomena, but with an emphasis on acquiring fundamental data rather than demonstrating new energy solutions (improves impartiality of findings).

Another approach to collecting and culling a suite of relevant proposals is through a sufficiently funded research solicitation. Procedures for conducting such a solicitation, inducing the ranking criteria, are already developed and refined from the NASA Breakthrough Propulsion Physics Project and are detailed in Chapter 22 of *FPS*.

## VII. CONCLUDING REMARKS

In the 1990's the subject of faster-than-light interstellar travel matured from mere science fiction into scientific discourse. By 2009, those and a host of other approaches toward breakthrough spaceflight matured to where the critical make-break issues are linked to unanswered questions in physics. Now, with those connections established, systematic rigorous research can commence.

Given the anomalies of deep space probe trajectories, Dark Matter, Dark Energy, and others, it is clear that further physics discoveries await. What is not clear is if these discoveries will also reveal new methods to traverse interstellar distances more effectively than rockets or light sails.

Progress is not made by conceding defeat. With a combination of risk-taking vision and impartial rigor, useful, reliable results will accumulate. Ad astra, incrementis (to the stars in steps, where each step is greater than before).